\documentclass[prl, twocolumn]{revtex4-2} 

\usepackage{graphicx}  
\usepackage{dcolumn}   
\usepackage{bm}        
\usepackage{amssymb}   
\usepackage{amsmath} 
\usepackage{color}
\usepackage{natbib}
\usepackage[utf8]{inputenc}
\usepackage[T1]{fontenc}
\usepackage[french,english]{babel}
\usepackage{filecontents}

\begin{document}

\title{Propagation of epidemics in a polarized society: impact of clustering among unvaccinated individuals}

\author{Ixandra Achitouv$^{1}$ }
\email{i.achitouv@imperial.ac.uk}
\affiliation{$^1$Department of Mathematics, Imperial College London}

\date{\today}

\begin{abstract} 

Polarization of opinions about vaccination can have a negative impact on pandemic control. In this work we quantify this negative impact for the transmission of COVID-19, using an agent based simulation in an heterogeneous population with multi-type networks, representing different types of social interactions. We show that the clustering of unvaccinated individuals, associated with polarization of opinion, can lead to significant differences in the evolution of the pandemic compared to deterministic model predictions. Under our realistic baseline scenario these differences are a $33\%$ increase of the effective reproduction number, a $157\%$ increase of  infections at the peak and a $30\%$ increase in the final cumulative attack rate. 
\end{abstract}

\maketitle

\section*{Introduction}
\noindent Social polarization is a segregation within a society resulting in groups at the extremities of the social hierarchy while shrinking groups around its middle \cite{1wiki}. It is an increasing area of research within developed economies \cite{3wiki}. Several factors such as income inequality and social media have been shown to increase polarization \cite{6wiki}. Indeed, social media tend to help cluster of acquaintances into homophilous circles, hence beeing exposed to news that is biased by its user’s choice \cite{7wiki}. In \cite{Salathe} the author shows that the risk of an influenza outbreak from vaccines hesitancy strongly increases when there is clustering of unvaccinated individuals. However, a direct impact of social polarization which have not been investigated to the best of our knowledge, is its impact on a global pandemic such as the COVID-19. In fact, Public Health measures often rely on scientific models such as \cite{SIRcov} or to estimate for instance herd immunity or forecast on intensive care unit. These deterministic models usually neglect the impact of social polarization by assuming that vaccines are homogeneously distributed in the population. However several studies show that vaccine acceptance and hesitancy correlate with a number of social economic factors and opinion. For instance in the US, vaccine hesitancy has been linked to political opinion, with 60$\%$ of Republican adults vs 91$\%$ of Democrats having received a first vaccine dose by December 2021 \cite{Allcott} ;~\cite{Yang}. In France, vaccine hesitancy has been linked to additional sociodemographic factors such as the level of education, the employed category (e.g. students, entrepreneurs,..) and standard of living. Within these categories, the vaccine coverage can be as different as 54$\%$ vs 88$\%$ for the bottom and top decile of the employed category \cite{Spire}.

\medskip
\noindent While social polarization, linked to vaccine hesitancy, reduces the overall vaccine coverage, it is unknown how it impacts the dynamic of the epidemic. Indeed, for a given number of  vaccine doses, the epidemic might be substantially different if unvaccinated individuals are clustered with each other rather than homogeneously distributed in the population. In this work we address this point by developing realistic simulations of the spread of SARS-Cov-2 when vaccines are allocated homogeneously, and by cluster, within a toy city of 100,000 inhabitants that have complex networks and a heterogeneous population.

\begin{figure}
\begin{center}
\includegraphics[scale=0.45]{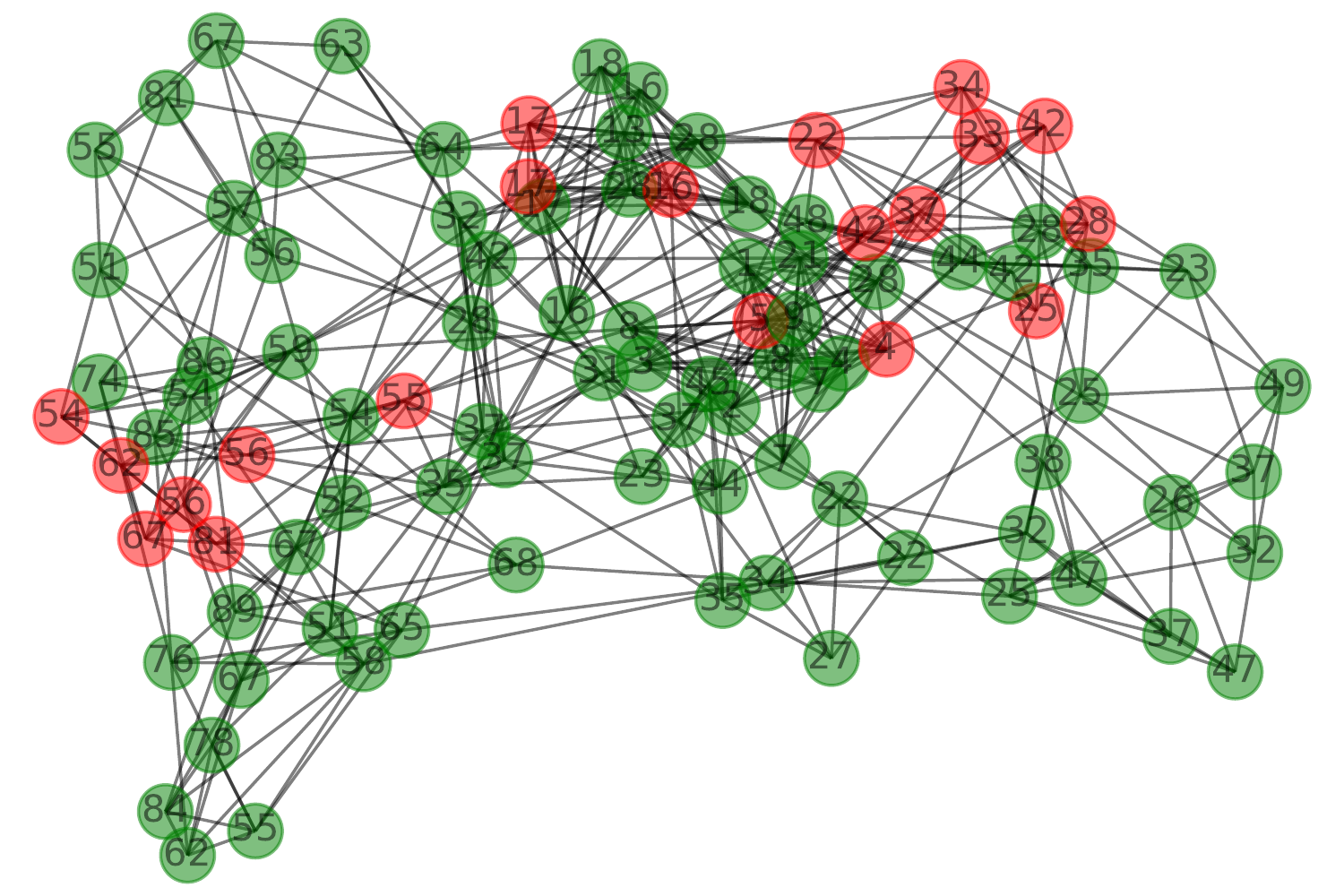}\\
\includegraphics[scale=0.45]{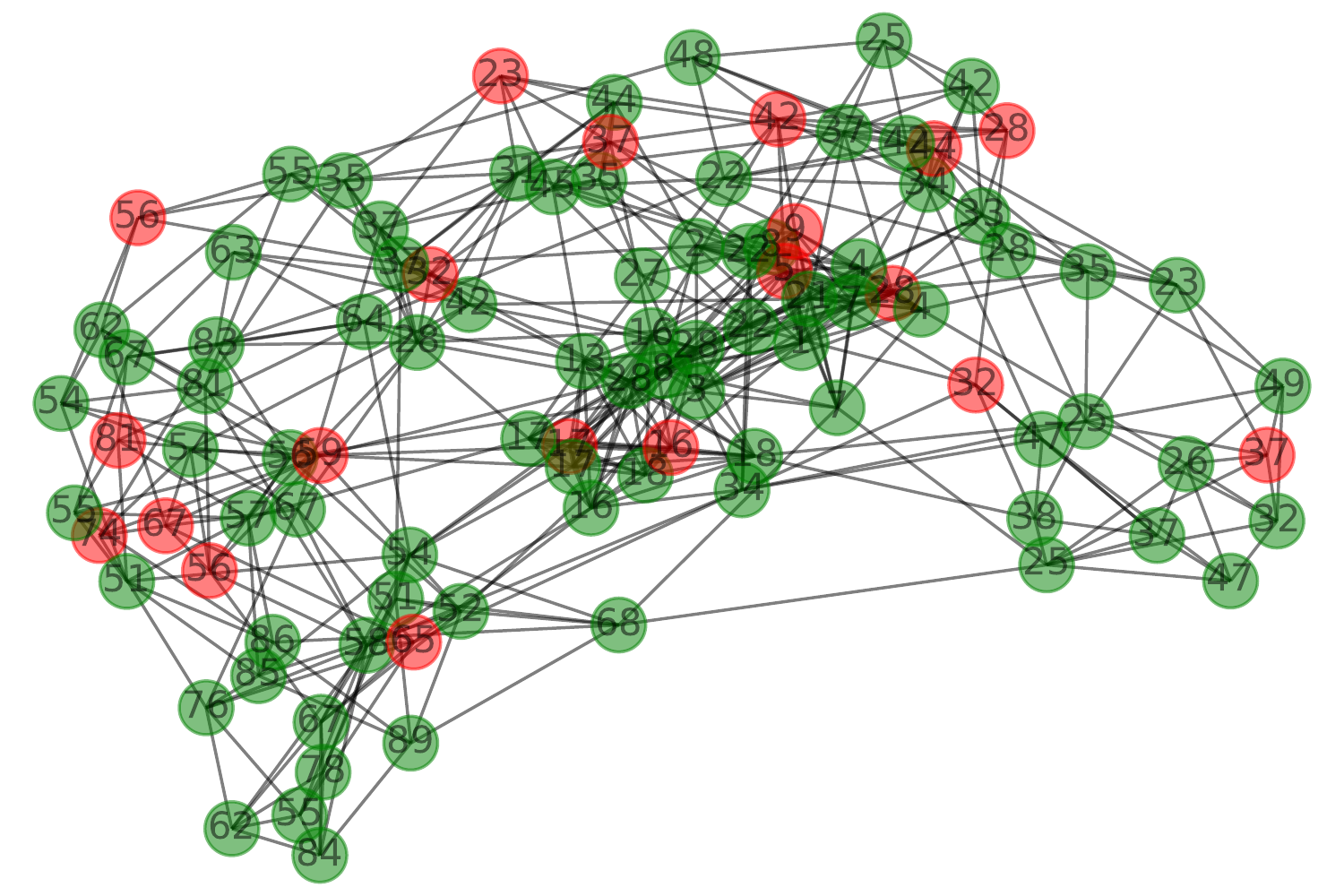}
\caption{Illustration of the interaction network between individuals: network with (top panel) and without (lower panel) clustering of the unvaccinated, for 100 individuals and 80$\%$ of vaccine coverage in each age categories (vaccinated/unvaccinated individuals in green/red). On each node we label the age of the agent.  
}\label{Fig1}
\end{center}
\end{figure}

\section*{Materials and methods}

\noindent We develop an agent based model with a structural network and infection characteristics of SARS-Cov-2 based on \cite{Hinch}. In our simulation the individuals are represented by the vertices and the edges represent the social contacts between them. The agents are structured by their age following the French demographics from the Insee data \cite{Insee} and are split into 4 categories that have specific mixing patterns: children, teenagers, adults and the elderly. For each category we model 3 types of daily interaction networks: i) between households members, ii) between members of an occupation network (workplace or school) which is stable over time and iii) incidental interactions drawn randomly on a daily basis. 

\noindent The household networks are static and are build from the French demographics and house sizes \cite{Insee}. Contact between all members of a house occurs daily.

\noindent The network structure for social interactions (workplace or school) is built using a small-world network, namely a Watts-Strogatz network \cite{Watts}. This is motivated by social experiments such as in ~\cite{Milgram}. Each individual has a fixed list of contact and every day we draw vertices with a fraction of his list to match the average number of contacts an individual of a given category has, using the social contact survey CoMix dataset for France \cite{CoMix}. To build the contact list, we also distinguish the category of the individual. For instance the list of recurring social interactions for children (e.g. school) is made with $80\%$ of children  and $20\%$ of adults as suggested in \cite{Hinch}.

\noindent The incidental interactions are drawn daily between individuals from a negative binomial distribution, an over-dispersed skewed distribution to account for "superspreaders". The parameters of this distribution depend of the age category of the individuals, to match the mean daily number of interactions outside of the social interactions from \cite{CoMix}.

\noindent Every day we loop over all infected individuals contacts and we compute a probability of infecting a non-infected individual from these lists. The probability of infection is the same as in Open ABM Covid19 (\cite{Hinch} and reference within) and is given by:

\begin{equation}
P(t,s_i,a_s,n)=1-\exp[-\lambda(t,s_i,a_s,n)], 
\end{equation}
with $\lambda$ is the rate at which the virus is transmitted in a single interaction:

\begin{equation}
\lambda(t,s_i,a_s,n)=\frac{R S_{a_s} A_{s_i} B_n }{\bar{I}}\int_{t-1}^{t} f_{\Gamma}(u;\mu_i;\sigma^{2}_{i}) du ,
\end{equation}

where $t$ is the time since infection, $s_i$ indicates the infector's symptom status (asymtomatic, mild, moderate/severe), $a_s$ is the age of the susceptible, $n$ is the type of network where the interaction occurred, $\bar{I}$ is the mean number of daily interactions, $f_{\Gamma}(u;\mu;\sigma^{2})$ is the probability density function of a gamma distribution with $\mu$ and $\sigma$ the mean and width of the infectiousness curve, $R$ scales the overall infection rate, $S_{a_s}$ is the scale-factor for the age of the susceptible, $A_{s_i}$ is the scale-factor for the infector being asymptomatic and $B_n$ is the scale-factor for the network on which the interaction occurred. The infection parameter values as a function of the age of the individuals can be found in \cite{Hinch} and references within. We checked that the results of our calibration are consistent with the age distribution of the positivity tests of COVID-19 in the French testing data (SI-DEP dataset \cite{SIDEP}) over the period 2020-09-01 until 2021-02-01 (before the vaccination of the population). 
 
 \begin{figure}
\begin{center}
\includegraphics[scale=0.45]{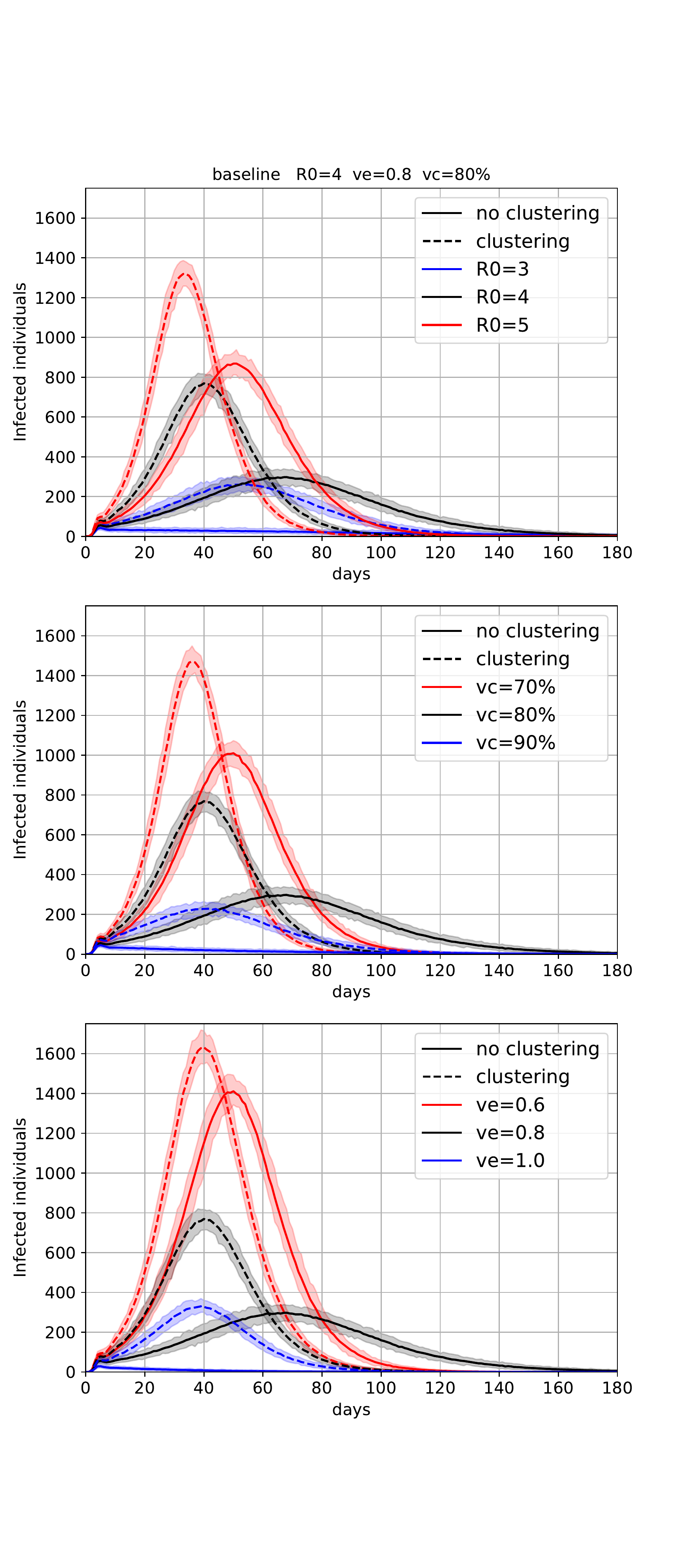}
\caption{Clustering impact on epidemic dynamics: daily number of infections over time, for different values of the basic reproduction number R0 (top panel), vaccination coverage (vc) (middle panel), and vaccine efficiency against infection (ve) (bottom panel). In each panel, the solid curves correspond to the scenario with no clustering of unvaccinated individuals while the dashed curves correspond to the scenario with clustering. Black curves are the same in all panels and correspond to the baseline model (R0=4, vc=80$\%$, ve=80$\%$).
}\label{Fig2a}
\end{center}
\end{figure}
\noindent  In order to characterize the degree of clustering of unvaccinated individuals we consider the assortativity which measures the tendency for vertices in networks to be connected to other vertices that are like (or unlike) them in some way \cite{Newman}. The assortativity is defined as
\begin{equation}
r=\frac{Tr \; \textbf{e}-\lVert \textbf{e}^2 \rVert}{1-\lVert \textbf{e}^2 \rVert},
\end{equation}

where $\textbf{e}$ is a 2x2 matrix whose elements are $<e_{i,j}>$ (the average is over all individuals) and the indices $i,j$ can have the values ${0,1}$ for unvaccinated and vaccinated individuals respectively. Thus $e_{1,1}$ corresponds to the fraction of connections a vaccinated agent has with other vaccinated agents, $e_{1,0}$ the fraction of connections a vaccinated agent has with unvaccinated agents and $e_{0,0}$ the fraction an unvaccinated agent has with other unvaccinated agents. For an undirected network, $\textbf{e}$ is symmetric. For homogeneous networks (where each agent are identical), $-1\leq r \leq 1$ where the limits $r=\left\lbrace -1,0,1\right\rbrace $ correspond to completely anti-correlated, random, and completely correlated groups. 

\noindent In what follows, we consider two values for the degree of clustering. The first case is $r=0$, which corresponds to no-clustering when the vaccines are allocated randomly across the population (what is usually assumed in mathematical models assessing how vaccines can mitigate the pandemic). The second case is $r=0.7$, which corresponds to the maximum assortativity value we could reach within the constraints of our network, and which corresponds to a model where unvaccinated individuals are clustered. To reach $r=0.7$ we randomly select one individual and start vaccinating all of its contacts while the number of vaccines in the age category of the contact is not reached. Then we allocate the vaccines to the contact of the contact etc. until there are no more vaccines to allocate. Finally we select individuals whose values of $e_{ij}$ with $i\neq j$ is larger than the mean, and we switch their vaccine status with someone of the same age as long as $r$ increases. We repeat this procedure for several random initial infected individuals and select the largest configuration that corresponds to the largest value of $r$. In our results $r=0.7$ is labeled as the clustering case. Fig.\ref{Fig1} shows a sample of our interaction network for 100 nodes, and illustrates how vaccines are distributed in the population in these two cases.

\section*{Results and Discussion}

\noindent To evaluate the impact of social hesitancy on vaccination, we run several simulations with clustering and no clustering of the vaccinated individuals. Our baseline scenario corresponds to 80$\%$ of vaccinated coverage (vc) in each age group, with a vaccine efficiency (ve) of 80$\%$ against infection and of 50$\%$ against transmission when infected. The basic reproduction number for the baseline scenario is $R_0=4$. For each scenario we run 100 simulations in order to estimate the uncertainty around the median value of daily infections. In Fig.\ref{Fig2a}, we show the daily number of infected individuals. For a fixed number of vaccine doses, the epidemic properties are significantly different when unvaccinated individuals are clustered than when they are homogeneously distributed in the interaction network. Sensitivity analysis also show that the clustering of the unvaccinated has a larger impact on epidemic dynamics for lower values of R0 (top panels), higher vaccine coverage (middle panels) and higher vaccine efficiency (bottom panels). For the baseline scenario (black curves), we have for the clustering (dashed curves) vs no clustering (solid curves) cases an effective reproduction number $R_{\rm eff}=1.6 \; \rm versius \; 1.2$ respectively. This leads to a faster growing epidemic for the clustering scenario, reaching its peak at $t=40 \; \rm versius \; 64$ days with a peak of $770 \; \rm versius \; 300$ infections for clustering vs no clustering (a 157$\%$ higher peak), and a total fraction of infected individuals of $23 \% \; \rm versius \; 30\%$ at the end of the simulation. In addition, we find that the epidemic goes extinct in the no-clustering network when the baseline scenario if changed assuming $R_0=3$ or $vc=90\%$ or $ve=1$ (top, middle, bottom panels and blue solid curve) which is not the case when there is clustering (dashed blue curves). This highlights, similarly to what was found in \cite{Salathe} for influenza, that herd immunity is highly sensitive to social polarization. 

\begin{figure}
\begin{center}
\includegraphics[scale=0.45]{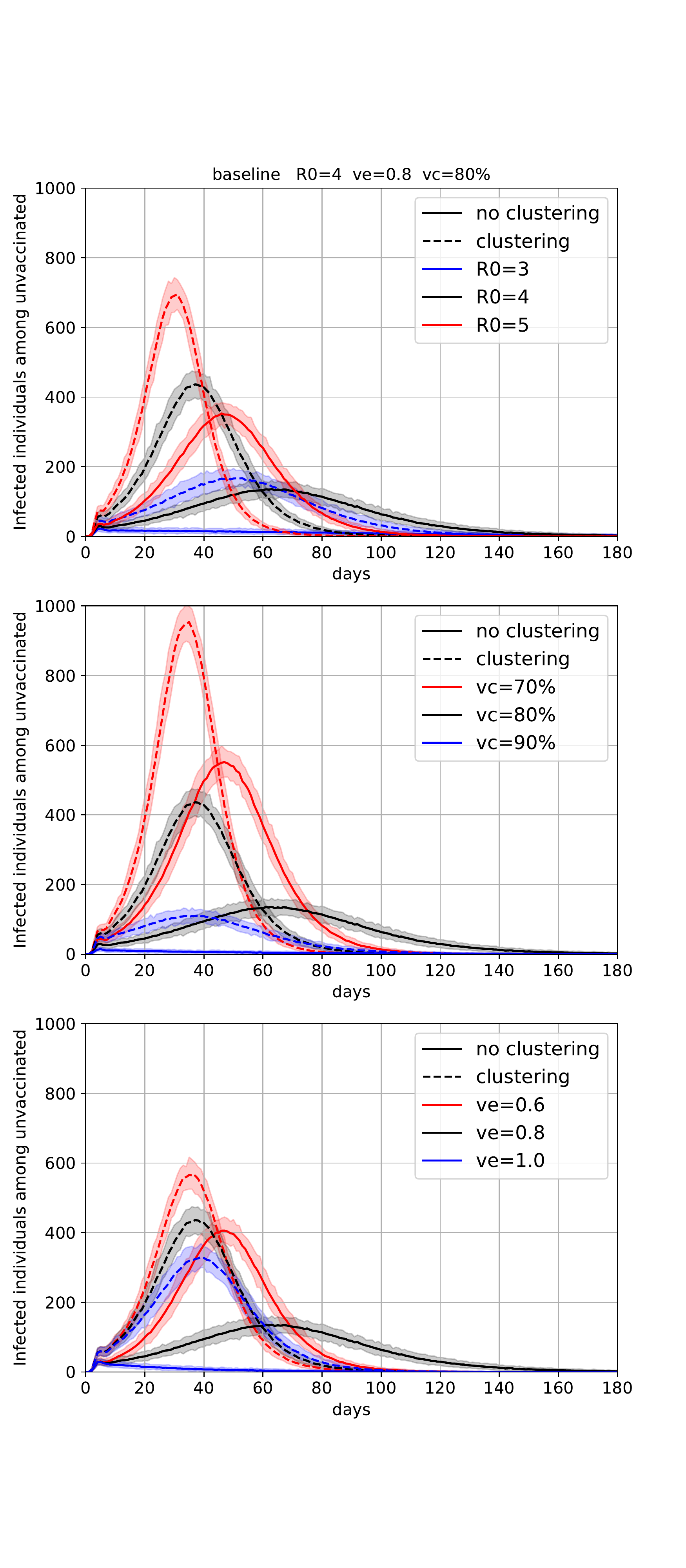}
\caption{Clustering impact on epidemic dynamics: daily number of infections over time among unvaccinated individuals, for different values of the basic reproduction number R0 (top panel), vaccination coverage (vc) (middle panel), and vaccine efficiency against infection (ve) (bottom panel). In each panel, the solid curves correspond to the scenario with no clustering of unvaccinated individuals while the dashed curves correspond to the scenario with clustering. Black curves are the same in all panels and correspond to the baseline model (R0=4, vc=80$\%$, ve=80$\%$).
}\label{Fig2b}
\end{center}
\end{figure}

\medskip
\noindent It is also interesting to quantify the impact of clustering among the population at risk i.e. the unvaccinated individuals. Indeed, we expect the non-linear impact of clustering to be higher within the unvaccinated population because in the extreme case ($r=1$ completely segregated unvaccinated versius vaccinated population), the diffusion of the virus won't be screened  by any vaccinated individuals. In Fig.\ref{Fig2b} we show the daily number of infections among the unvaccinated individuals for the same parameters as Fig.\ref{Fig2a}. For the baseline scenario (black curves), we report a peak of $136 \; \rm versius \; 436$ infections, thus a 221$\%$ higher peak and a total fraction of infected individuals of $10.2 \%$ versius $15.8\%$ for the no clustering vs clustering scenario respectively. This result is similar to a recent study by \cite{Fisman} using a simple Susceptible-Infectious-Recovered (SIR) compartmental model.

\section*{Conclusion}

\noindent Overall, the polarization of opinions about vaccination in many countries has reduced the overall vaccine coverage, thereby negatively impacting pandemic control. In addition, this polarization {has likely generated important clusters of unvaccinated individuals}. In this paper, we report that such clustering further significantly {reduces the impact of a vaccination campaign on disease spread} by (i) accelerating disease spread in the community and (ii) increasing the final cumulative attack rate.  This impact is accentuated in the unvaccinated population which is of particular concern since this population lacks protection.

\medskip
\noindent Several article have recently pointed out how heterogeneities shape epidemics dynamics \cite{Grossmann}, ~\cite{Nielsen} and that the widely-used class of deterministic mathematical models known as ODE models cannot accurately capture population heterogeneity \cite{Kuhl}, ~\cite{Holmdahl}. The clustering of opinions is one of these heterogeneities, rarely accounted for, that could potentially degrade model projections. This highlights the importance of better understanding and modelling the polarization within societies to control the course of a global epidemic.

\section*{Acknowledgments}
\noindent I would like to thank P. Bosetti and N. Hoz\'e for the fruitful discussions on assortativity, S. Cauchemez and L. Opatowski for their useful discussions on general topics related to COVID-19.

\medskip

\bibliographystyle{plain}
\bibliography{Ref.bib}

\end{document}